\documentstyle[pra,aps,multicol]{revtex} 

\input{epsf}

\tighten  

\begin{document}
\draft
\title{Heisenberg picture operators in the stochastic wave  function approach
to open quantum systems.}

\author{ H. P. Breuer, B. Kappler and F. Petruccione }
\address{
Albert-Ludwigs-Universit\"at, Fakult\"at f\"ur Physik, \\
Hermann-Herder Stra{\ss}e 3, D--79104 Freiburg im Breisgau,
Federal Republic of Germany}
\date{\today}
\maketitle


\begin{abstract}
A fast simulation algorithm for the calculation of multitime correlation
functions of open quantum systems is presented. It is demonstrated that any
stochastic process which ``unravels'' the quantum Master equation can be 
used for the calculation of matrix elements of reduced Heisenberg picture 
operators, and thus for the calculation of multitime correlation functions,
by extending the stochastic process to a doubled Hilbert space. The numerical 
performance of the stochastic simulation algorithm is
investigated by means of a standard example.

\end{abstract}

\begin{multicols}{2}
\narrowtext
\section{Introduction}

The state of an open quantum system is conventionally described through a 
reduced density matrix  $\rho(t)$ whose dynamics is given by a dissipative
equation of motion -- the quantum Master equation.
From a numerical point of view this formalism has a major
drawback:  for a system whose state is described in a $N$-dimensional Hilbert
space ${\cal H}$, the quantum Master equation is a set of $N(N+1)/2$ coupled 
differential equations. Hence the numerical evaluation of the quantum Master 
equation is in practice not feasible for large systems \cite{MolmerOpt}.

This difficulty does not arise in the stochastic wave 
function approach to open systems 
\cite{MolmerPRL68,Carmichael,GardinerPRA46,ZollerPRA46,BP:QS3,BP:QS4,Gisin:92,Gisin:93}: Here, the state of an open quantum system
is described by a stochastic wave function $\psi(t)\in{\cal H}$, {\it i.e.},
by a $N-$dimensional state vector. The stochastic time evolution of 
$\psi(t)$ is
either defined through a stochastic Schr\"odinger equation 
\cite{Gisin:92,Gisin:93} (which is a 
stochastic differential equation) or alternatively through a conditional 
transition probability $T[\psi,t|\psi_0,t_0]$ \cite{BP:QS3,BP:QS4}, 
which is the probability density
of finding the system in the state $\psi$ at time $t\ge t_0$ {\it under
the condition} that the system is in the state $\psi_0$ at time $t_0$.
The connection to the density matrix formalism is made through the 
relation
\begin{eqnarray}
  \label{rho_P_eq}
  \rho(t)&=&\int D\psi D\psi^*\int D\psi_0 D\psi_0^* 
  \;|\psi\rangle\langle\psi|\; \nonumber\\
  &&\times T[\psi,t|\psi_0,t_0]P[\psi_0,t_0],
\end{eqnarray}
where $P[\psi_0,t_0]$ is the probability density of an ensemble of
normalized pure states characterizing some initial density matrix
$\rho_0$ and $D\psi D\psi^*$ is the Hilbert space volume element
\cite{BP:QS3,BP:QS4}. The integrals extend over the Hilbert space
${\cal H}$. This relation ensures that one-time expectation values of
any system operators are calculated correctly.  Note that this
condition alone does not uniquely specify a stochastic
process. Diffusion type stochastic processes \cite{Gisin:92,Gisin:93}
as well as piecewise deterministic jump processes
\cite{MolmerPRL68,Carmichael,GardinerPRA46,ZollerPRA46,BP:QS3,BP:QS4}
have been proposed in the literature. A unique stochastic process can
only be derived by making further assumptions such as specifying a
certain measurement scheme \cite{WisemanMilburnPRA93,WisemanPRA93,BP:QS8}.

Especially in quantum optical systems one-time expectation values of system 
observables are not the only measurable quantities: for example,
the spectrum of fluorescence of a two level system is the Fourier transform
of the two-time correlation function $\langle\!\langle 
\sigma^+(\tau)\sigma^-\rangle\!\rangle_{\rm  s}$ in the stationary state, 
where $\sigma^\pm$ denote the pseudo spin operators of the system.
 Thus, for a complete description of open quantum systems it is 
necessary to introduce Heisenberg picture operators. In the density matrix
formalism this concept is well understood \cite{Alicki}: Consider the
quantum Master equation 
\begin{equation}
  \label{QME_eq}
  \dot\rho(t)=L(t)\rho(t),
\end{equation}
where the super-operator $L(t)$ is defined as
\begin{eqnarray}
  \label{sup_eq}
  L(t)\rho(t)&=&-i\left[H(t), \rho(t)\right]\\
  &&+\frac{1}{2}\sum_i\gamma_i\Big\{
  2J_i\rho(t) J_i^\dagger-J_i^\dagger J_i\rho(t)-\rho(t)J_i^\dagger
  J_i\Big\}.\nonumber
\end{eqnarray}
The operator $H(t)$ is essentially the Hamiltonian of the
isolated system which contributes to the 
coherent part of the dynamics, and  the rates $\gamma_i$ and the 
Lindblad operators $J_i$ 
describe the dissipative coupling of the system to its environment through
the $i$-th decay channel. The solution of eq.~(\ref{QME_eq}) with respect to 
the initial condition $\rho(t_0)=\rho_0$ can be expressed
for $t\ge t_0$ in terms of the propagation super-operator $V(t,t_0)$ as
\cite{GardinerQN}
\begin{equation}
  \label{sol_eq}
  \rho(t)=V(t,t_0)\rho_0,
\end{equation}
where $V(t,t_0)$ is the solution of the differential equation
\begin{equation}
  \label{prop_diff_eq}
  \frac{d}{d t}V(t,t_0)=L(t)V(t,t_0),
\end{equation}
with the initial condition $V(t_0,t_0)=I$.
For an arbitrary Schr\"odinger system operator $A$ the 
matrix elements of the {\it reduced} Heisenberg picture operator are defined 
as 
\begin{eqnarray}
  \label{heis_def_eq}
  A_t(\phi_0,\psi_0) &\equiv& \langle\phi_0,t_0|A(t)|\psi_0,t_0\rangle\nonumber\\
  &=&\mbox{Tr}_{\rm sys}\Big\{AV(t,t_0)|\psi_0\rangle\langle\phi_0|\Big\}.
\end{eqnarray}
Eq.~(\ref{heis_def_eq}) can be interpreted in the following way: for the 
calculation of the matrix element $A_t(\phi_0,\psi_0)$ start with 
the initial ``density matrix'' $|\psi_0\rangle\langle \phi_0|$ and propagate
 it up to the time $t$. Then calculate the expectation value of
$A$ with respect to the propagated ``density matrix''. However,  since  
$|\psi_0\rangle\langle \phi_0|$ is
in general not a positive matrix and thus not a true density matrix, it can 
not be characterized by a probability density $P[\psi_0,t_0]$ of normalized
pure states in ${\cal H}$ (cf. eq.~(\ref{rho_P_eq}). Hence a direct application of the stochastic wave
function approach to the calculation of Heisenberg picture operators is 
not possible. 

\section{Heisenberg picture operators in the stochastic wave function approach}

In a closed system where the time evolution of states is 
given through the unitary propagator $U(t,t_0)$ we
can calculate arbitrary matrix elements $A_t(\phi_0,\psi_0)$ of a Heisenberg 
operator $A(t)$ in the following
way (cf. fig.~\ref{heis_calc}): propagate $\phi_0$ and $\psi_0$ to obtain 
$\phi=U(t,t_0)\phi_0$ and $\psi=U(t,t_0)\psi_0$, respectively and then 
evaluate the scalar product $\langle\phi|A|\psi\rangle$. This method
is easily generalized to the calculation of matrix elements of 
a {\it reduced} Heisenberg picture operator, {\it i.e.}, to open systems: 
Instead of propagating the state vectors $\phi_0\in{\cal H}$ and 
$\psi_0\in{\cal H}$ separately, we can construct a stochastic process
in the {\it doubled} Hilbert space $\widetilde{\cal H}={\cal H}
\oplus{\cal H}$ which propagates the normalized {\it pair} of state vectors
$\theta_0=(\phi_0,\psi_0)^{\rm T}/\sqrt{2}\in\widetilde{\cal H}$ simultaneously
in such a way that the following condition holds:
\begin{equation}
  \label{mat_exp_eq}
  A_t(\phi_0,\psi_0)=2\int D\theta D\theta^* \langle\phi|A|\psi\rangle
  \widetilde{T}[\theta,t|\theta_0,t_0],
\end{equation}
where $\theta=(\phi,\psi)^{\rm T}$ and we introduced the conditional
transition probability $\widetilde{T}$ for the stochastic process in
the doubled Hilbert space $\widetilde{\cal H}$.  Throughout this
letter, the superscript T denotes the transpose of a vector.  This
condition states that matrix elements of arbitrary Heisenberg
operators are calculated correctly.  It is important to note that
eq.~(\ref{mat_exp_eq}) alone does not specify the stochastic time
evolution in the doubled Hilbert space uniquely. In fact, each
stochastic process which can be used to simulate the quantum Master
equation (\ref{QME_eq}) can be extended to the doubled Hilbert space
and used for the calculation of the matrix elements of arbitrary
Heisenberg picture operators. We will first demonstrate this for the
piecewise deterministic jump process proposed in
\cite{MolmerPRL68,Carmichael,GardinerPRA46,ZollerPRA46,BP:QS3,BP:QS4}
and then generalize the result.  A derivation of the simulation
algorithm for the stochastic time evolution in the doubled Hilbert
space which is based on a microscopic system--reservoir model can be
found in Ref.~\cite{BP:QS12}.

For the piecewise deterministic jump process the simulation algorithm reads as
follows:
1) Start with the normalized state $\theta_0=(\phi_0,\psi_0)^{\rm T}/\sqrt{2}$
at $t_0$.
2) Draw a random number $\eta_1$ from a uniform distribution on $[0,1]$;
this random number will determine the time of the first jump.
3) Propagate $\theta_0$ according to the Schr\"odinger-type equation
\begin{equation}
  \label{det_eq}
  i\frac{d}{ds}\hat\theta(s)=\widetilde{H}_{\rm eff}(s)\hat\theta(s)
\end{equation}
where the extensions of the Hamiltonian $H$ and Lindblad operators $J_i$ 
to the doubled Hilbert space are defined as 
\begin{equation}
\label{H_J_ext_eq}
  \widetilde{H}(s)=
  \left(\begin{array}{cc}
  H(s)&0\\
  0&H(s)\end{array}\right),\quad
  \widetilde{J}_i=
  \left(\begin{array}{cc}
  J_i&0\\
  0&J_i\end{array}\right),
\end{equation}
and the non-Hermitian effective Hamiltonian is defined as
\begin{equation}
  \label{H_eff_eq}
  \widetilde{H}_{\rm  eff}(s)=\widetilde{H}(s)-\frac{i}{2}\sum_i \gamma_i\widetilde{J}_i^\dagger \widetilde{J}_i.
\end{equation}
4) The time $T_1$ of the first jump is determined by the condition
\begin{equation}
  \label{jump_cond_eq}
  \eta_1=\|\hat\theta(T_1)\|^2.
\end{equation}
5) Select a particular type of jump with probability  
$\gamma_i w_i/\sum_i\gamma_i w_i$, where 
$w_i=\|\widetilde{J}_i\hat\theta(T_1)\|^2$. 
The state of the system immediately after the first jump is given by
\begin{equation}
  \label{aft_jump_eq}
  \theta(T_1)=\widetilde{J}_i\hat\theta(T_1)
  /\|\widetilde{J}_i\hat\theta(T_1)\|.
\end{equation}
6) Draw a second random number $\eta_2$ to determine the time $T_2$ of the 
next jump and propagate $\theta(T_1)$ according to the differential
equation (\ref{det_eq}) and so on until $s=t$.
7) The state of the system at time $t$ is given by 
\begin{equation}
  \label{final_eq}
  \theta(t)\equiv(\phi(t),\psi(t))^{\rm T}=\hat\theta(t)/\|\hat\theta(t)\|.
\end{equation}

The matrix elements of the reduced Heisenberg picture operator 
$A(t)$ are then obtained by computing
\begin{equation}
  \label{Heis_calc_eq}
  A_t(\phi_0,\psi_0)=2\Big\langle\!\Big\langle\langle\phi(t)|A|
  \psi(t)\rangle\Big\rangle\!\Big\rangle,
\end{equation}
where the angular brackets $\langle\!\langle\cdots\rangle\!\rangle$ denote
the average over the realizations of the stochastic process. 

In order to show that this algorithm leads to the correct result, we
introduce the density matrix 
\begin{equation}
  \label{rho_tilde_eq}
  \widetilde{\rho}(t)=\left(
  \begin{array}{cc}
  \widetilde{\rho}_{11}(t)&\widetilde{\rho}_{12}(t)\\
  \widetilde{\rho}_{21}(t)&\widetilde{\rho}_{22}(t)\end{array}\right)
\end{equation}
on the doubled Hilbert space
$\widetilde{\cal H}$ which is a solution of the extended quantum Master 
equation
\begin{eqnarray}
  \label{ext_QME_eq}
  \dot{\widetilde\rho}(t)&=&-i\left[\widetilde{H}(t),
    \widetilde\rho(t)\right]\\
  &&+\frac{1}{2}\sum_i\gamma_i\Big\{
  2\widetilde{J}_i\widetilde\rho(t) \widetilde{J}_i^\dagger-
  \widetilde{J}_i^\dagger \widetilde{J}_i\widetilde\rho(t)-
  \widetilde\rho(t) \widetilde{J}_i^\dagger \widetilde{J}_i\Big\},\nonumber
\end{eqnarray}
with the initial condition
\begin{equation}
  \label{ext_init_eq}
  \widetilde\rho(t_0)=|\theta_0\rangle\langle\theta_0|\equiv\frac{1}{2}\left(
  \begin{array}{cc}
  |\phi_0\rangle\langle\phi_0|&|\phi_0\rangle\langle\psi_0|\\
  |\psi_0\rangle\langle\phi_0|&|\psi_0\rangle\langle\psi_0|\end{array}\right).
\end{equation}
By definition of the extended operators $\widetilde{H}$ and 
$\widetilde{J}_i$ each component $\widetilde{\rho}_{ij}(t)$ which is
an operator on ${\cal H}$ is a solution
of the original quantum Master equation (\ref{QME_eq}). Since 
$\widetilde{\rho}_{21}(t)$  is a solution of eq.~(\ref{QME_eq}) with the
initial condition $\widetilde{\rho}_{21}(t_0)=|\psi_0\rangle\langle\phi_0|/2$
the matrix elements $A_t(\phi_0,\psi_0)$ of a reduced Heisenberg
picture operator $A(t)$ can be written as (cf. eq.~(\ref{heis_def_eq}))
\begin{equation}
  \label{Heis_ext_eq}
  A_t(\phi_0,\psi_0)=
  2\mbox{Tr}_{\rm sys}\Big\{A\widetilde{\rho}_{21}(t)\Big\}.
\end{equation}
Now consider a particular ``unraveling'' of the {\it extended}
 quantum Master equation 
(\ref{ext_QME_eq}) which is characterized by a conditional transition 
probability $\widetilde T[\theta,t|\theta_0,t_0]$ in the doubled Hilbert 
space. For the density matrix $\widetilde{\rho}(t)$ we then obtain in analogy
to eq.~(\ref{rho_P_eq})
\begin{equation}
  \label{rho_tilde_T_eq}
  \widetilde{\rho}(t)=\int D\theta D\theta^*\;|\theta\rangle\langle\theta|\;
  \widetilde T[\theta,t|\theta_0,t_0],
\end{equation}
and hence for $\widetilde{\rho}_{21}(t)$ 
\begin{equation}
  \label{rho_21_tilde_T_eq}
  \widetilde{\rho}_{21}(t)=\int D\theta D\theta^*\;|\psi\rangle\langle\phi|\;
  \widetilde T[\theta,t|\theta_0,t_0],
\end{equation}
where $\theta=(\phi,\psi)^{\rm T}$. By inserting eq.~(\ref{rho_21_tilde_T_eq})
into eq.~(\ref{Heis_ext_eq}) we recover eq.~(\ref{mat_exp_eq}).
Thus we have shown that matrix elements of reduced Heisenberg picture
operators are calculated correctly ({\it i.e.}, eq.~(\ref{mat_exp_eq}) holds)
if the stochastic process in the doubled Hilbert space can be used
to simulate the extended quantum Master equation (\ref{ext_QME_eq}). 
Since this is the case for the simulation algorithm presented above we have 
completed the proof.

It is important to note that the above proof does not rely on a specific
``unraveling''  of the quantum Master equation (\ref{ext_QME_eq}). On the
contrary, it is valid for any stochastic process the covariance matrix
(see eq.~(\ref{rho_tilde_T_eq})) of which is governed by 
eq.~(\ref{ext_QME_eq}).

\section{Multitime correlation functions}

The simulation algorithm in the doubled Hilbert space
can also be used for the calculation of multitime correlation functions.
Consider for example the two-time correlation function 
\begin{equation}
  \label{corr_eq}
  g(\phi_0,t_1,t_2)=\langle\phi_0,t_0|A(t_2)B(t_1)|\phi_0,t_0\rangle,
\end{equation}
where $t_1\le t_2$. Here, the stochastic simulation algorithm would read as 
follows: 1) Start in the state $\phi_0$ at time $t_0$ and use the stochastic 
time evolution in the Hilbert space ${\cal H}$ to obtain the stochastic 
wave function $\phi(t_1)$.
2) Propagate the state 
\begin{equation}
  \label{corr_sim_eq}
  \theta(t_1)=(\phi(t_1),B\phi(t_1))^{\rm T}/\|(\phi(t_1),B\phi(t_1))\|
\end{equation}
using the stochastic time evolution in the 
doubled Hilbert space to obtain the state vector $\theta(t_2)=
(\phi(t_2),\psi(t_2))^{\rm T}$.
The multitime correlation function is then obtained 
by computing 
\begin{equation}
  \label{sim_corr_2_eq}
  g(\phi_0,t_1,t_2)=\Big\langle\!\Big\langle 
  \big\|\big(\phi(t_1),B\phi(t_1)\big)\big\|^2
  \langle\phi(t_2)|A|\psi(t_2)\rangle\Big\rangle\!\Big\rangle.
\end{equation}

The generalization of this scheme to the calculation of arbitrary 
time-ordered multitime correlation functions of the form
\begin{eqnarray}
\label{gen_corr_eq}
\lefteqn{g(\phi_0,t_0,t_1,...,t_n,s_1,...,s_m)}\\
&=&\langle{\phi_0,t_0}|A_1(t_1)\cdots 
A_n(t_n)B_m(s_m)\cdots B_1(s_1)|{\phi_0,t_0}\rangle,\nonumber
\end{eqnarray}
where $t_0\le\cdots\le t_n$, and $t_0\le s_1\le\cdots\le s_m$, and
$A_i$ and $B_i$ are arbitrary system operators is straightforward: 
Order the set of times $\{t_1,\cdots t_n,s_1,\cdots s_m\}$ and rename
 them $r_i$ such
that $r_1<\cdots<r_{q}$ where $q$ is the number of distinct time points.
 Then define a set of Schr\"odinger operators $F_l$ and $G_l$ as
\begin{equation}
  \label{F_G_eq}
 \left\{\begin{array}{cl}
     F_l = A_i^\dagger, G_l = I,
       & \mbox{if $r_l=t_i\neq s_j$ for some $i$ and all $j$,}\\
     G_l = I, F_l = B_j,
       & \mbox{if $r_l= s_j\neq t_i$ for some $j$ and all $i$,}\\
     G_l = A_i^\dagger, F_l = B_j,
       & \mbox{if $r_l=t_i= s_j$ for some $i$ and $j$}.
\end{array}\right.
\end{equation}
The multitime correlation function
$g(\phi_0,t_0,t_1,...,$ $t_n,s_1,...,s_m)$ is then obtained in the
following way: 

1) Start with the state $\phi_0$ at time $t_0$  and
propagate it up to the time $r_1$ to obtain $\phi(r_1)$. 

2) Propagate the state 
\begin{equation}
  \label{gen_theta_eq}
  \theta(r_1)= \left(F_1\phi(r_1),G_1\phi(r_1)\right)^{\rm T}/ \|
  (F_1\phi(r_1),G_1\phi(r_1))\| 
\end{equation}
to obtain $\theta(r_2)=(\phi(r_2),\psi(r_2))^{\rm T}$. 

3) Jump to the state 
\begin{equation}
  \label{gen_thet_2_eq}
  \theta(r_2)= \left(F_2\phi(r_2),G_2\psi(r_2)\right)^{\rm T}/ \|
  (F_2\phi(r_2),G_2\psi(r_2))\| 
\end{equation}
and propagate it up to $r_3$ and so
on. $g(\phi_0,t_0,t_1,...,t_n,s_1,...,s_m)$ is then given by 
\begin{eqnarray}
  \label{g_gen_eq}
  \lefteqn{g(\phi_0,t_0,t_1,...,t_n,s_1,...,s_m)=}\\
  &&\Big\langle\!\Big\langle 
  \big\|\big(F_1\phi(r_1),G_1\phi(r_1)\big)\big\|^2
 \big\|\big(F_2\phi(r_2),G_2\psi(r_2)\big)\big\|^2\cdots\nonumber\\
  &&\times\big\|\big(F_{q-1}\phi(r_{q-1}),G_{q-1}\psi(r_{q-1})\big)\big\|^2 
  \langle\phi(r_{q})|F_{q}^\dagger
  G_{q}|\psi((r_{q})\rangle\Big\rangle\!\Big\rangle.\nonumber  
\end{eqnarray}
It is important to note, that also for higher order correlation
functions, we only have to propagate two state vectors. 

Finally, let us remark that the choice of the initial
condition~(\ref{sim_corr_2_eq}) (or (\ref{gen_theta_eq}), respectively)
is not unique. We can also multiply the operator $B$ by a constant
$\varepsilon$ and define the state vector $\theta_\varepsilon(t_1)$
as
\begin{equation}
  \label{theta_eps_eq}
  \theta_\varepsilon(t_1)=(\phi(t_1),\varepsilon B\phi(t_1))^{\rm
  T}/\|(\phi(t_1),\varepsilon B\phi(t_1))\|
\end{equation}
and accordingly the correlation function $g(\phi_0,t_1,t_2)$ as
\begin{eqnarray}
  \label{corr_eps_eq}
  \lefteqn{g(\phi_0,t_1,t_2)=}\\
  &&\frac{1}{\varepsilon}\Big\langle\!\Big\langle 
  \big\|\big(\phi(t_1),\varepsilon B\phi(t_1)\big)\big\|^2
  \langle\phi_\varepsilon(t_2)|A|
  \psi_\varepsilon(t_2)\rangle\Big\rangle\!\Big\rangle,\nonumber
\end{eqnarray}
where $\theta_\varepsilon(t_2)=(\phi_\varepsilon(t_2),
\psi_\varepsilon(t_2))^{\rm T}$ is obtained by propagating
$\theta_\varepsilon(t_1)$ according to the simulation algorithm in the
doubled Hilbert space. Again, the unnormalized deterministic motion is
governed by the equation of motion
\begin{eqnarray}
  \label{phi_psi_eq}
  i\frac{d}{dt}\hat{\phi}_\varepsilon(t)&=& H_{\rm
  eff}(t)\hat{\phi}_\varepsilon(t)\\
  i\frac{d}{dt}\hat{\psi}_\varepsilon(t)&=& H_{\rm
  eff}(t)\hat{\psi}_\varepsilon(t)
\end{eqnarray}
but in the limit $\varepsilon\rightarrow 0$, we find 
\begin{eqnarray}
  \label{norm_w_eq}
  \|\hat\theta_\varepsilon(t)\|&\rightarrow\|&\hat\phi_\varepsilon(t)\|\\
  w_i=\|\widetilde{J}_i\hat\theta_\varepsilon(T)\|^2&\rightarrow&
  \|J_i\hat\phi_\varepsilon(T)\|^2,
\end{eqnarray}
and hence the jumps of the trajectory $\theta_\varepsilon(t)$ are
completely governed by the jumps of $\phi_\varepsilon(t)$, which
evolves according to the ``usual'' stochastic time evolution in ${\cal
  H}$ (cf. Eqs.~(\ref{jump_cond_eq}) and (\ref{aft_jump_eq})). In this
limit we obtain a procedure first proposed by Dum {\it et~al.} in
Ref. \cite{ZollerPRA46}, which is based on ``probing the system with
$\delta$ kicks'' (see Appendix D of Ref.\cite{ZollerPRA46}). For
further discussions of this method see for example the Refs. \cite{MartePRA,MartePRL,MolmerSemOpt}.

\section{Numerical results}

\begin{figure}[t]
  \epsfxsize\linewidth\epsffile{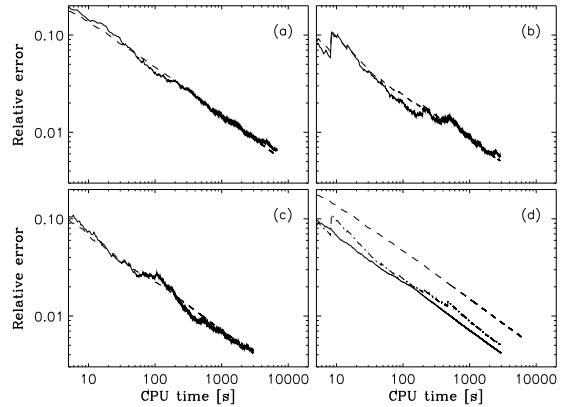}
    \caption{\label{num_per}
    Calculation of the first order correlation function
    $\langle\!\langle \sigma^+(\tau)\sigma^-\rangle\!\rangle_{\rm s}$
    for a coherently driven two level atom on resonance. This figure
    shows the relative error vs.~the CPU time in seconds for the
    simulation algorithms proposed in Ref. \protect\cite{MolmerPRL68} (a),
    Ref. \protect\cite{ZollerPRA46} (b), and for our algorithm in the
    doubled Hilbert space (c).  The solid lines represent the mean
    square deviation of the numerical solution from the exact solution
    and the dashed lines show the estimated standard deviation of the
    numerical solution. In Fig.~\ref{num_per} (d), we compare the
    estimated standard deviation for the three algorithms.}
\end{figure}

In order to investigate the numerical performance of our simulation
algorithm, we compare it with the method proposed by Dum {\it et~al.}
in Ref. \cite{ZollerPRA46} and with an alternative method proposed by
Dalibard {\it et~al.} which is based on a decomposition of the
stochastic trajectory into four sub-trajectories \cite{MolmerPRL68}.
Note that all procedures are fully consistent with the quantum
regression theorem \cite{GardinerQN,Walls} and hence lead to the same
result for the multitime correlation function. However, the numerical
performance of the algorithms is quite different. We demonstrate this
by means of a standard example of quantum optics -- the calculation of
the spectrum of resonance fluorescence of a two level system.  In
fig. \ref{num_per} (a) -- (c) we show the computational time necessary
to achieve a given accuracy (measured by the relative error of the
correlation function $\langle\!\langle
\sigma^+(\tau)\sigma^-\rangle\!\rangle_{\rm s}$ in the stationary
state) for a coherently driven two level atom with Rabi frequency
$\Omega=10\gamma$ obtained on a RS6000 workstation. The solid lines
represent the mean square deviation of the numerical solution from the
exact solution \cite{Mollow:spec} and the dashed lines show the mean
estimated standard deviation of the numerical solution.  Obviously,
the latter quantity provides for all algorithms a very good measure of
the accuracy of the numerical simulation. In fig. \ref{num_per} (d) we
compare the estimated standard deviation for the three algorithms.
Obviously, the numerical performance of the algorithms proposed by Dum
{\it et.~al.} and our algorithm is quite similar, although the
convergence of our algorithm is smoother. On the other hand, for a
given accuracy the stochastic simulation in the doubled Hilbert space
is by a factor of $3$ faster than the algorithm proposed in
\cite{MolmerPRL68}.  We expect this result to be even better for
higher order correlation functions since for a multitime correlation
function of the type of eq.~(\ref{gen_corr_eq}) one has to propagate
in general $4^{n+m-1}$ different state vectors in each realization
using the method of Castin {\it et.~al}, whereas in our approach it is
only necessary to propagate two state vectors.
 
Let us briefly summarize the main results of this letter: We have shown that
starting from a stochastic simulation algorithm for the quantum Master 
equation (\ref{QME_eq}) it is possible to obtain a fast simulation 
algorithm for the calculation of matrix elements of arbitrary Heisenberg 
picture operators and time-ordered multitime correlation functions by 
making the substitutions
\begin{equation}
  \label{repl_eq}
  \psi\in{\cal H}\longrightarrow \theta\in\widetilde{\cal H},\quad
  H\longrightarrow \widetilde{H},\quad
  J_i\longrightarrow \widetilde{J}_i,
\end{equation}
{\it i.e.}, we replace the stochastic wave function $\psi(t)$ by a stochastic 
wave function $\theta(t)$ in the doubled Hilbert
space and extend accordingly the operators $H$ and $J_i$ which are present
in the quantum Master equation to the doubled Hilbert space 
(cf. eq.~(\ref{H_J_ext_eq})). We emphasize that these 
replacements can be done for {\it any} ``unraveling''  of the quantum Master 
equation, {\it e.g.}, also for the quantum state diffusion model 
\cite{Gisin:92,Gisin:93}.
The resulting stochastic process in the doubled Hilbert space 
is then similar to a process first proposed by Gisin in \cite{Gisin:Heis}.
However, the latter process is only well defined, when the initial
states $\phi_0$ and $\psi_0$ are {\it non-orthogonal}, {\it i.e.}, if $\langle
\phi_0|\psi_0\rangle\ne 0$. This problem does not occur in the  
ansatz presented here.

\bibliographystyle{prsty}  

\parbox{\linewidth}{
\begin{figure}[t]
  \epsfxsize\linewidth\epsffile{fig1.eps}
    \caption{\label{heis_calc}
    Calculation of Heisenberg operator matrix elements: (a) for a 
    closed system and (b) for an open system.}
\end{figure}
}

\vspace*{70ex}
\end{multicols}
\end{document}